\documentclass[a4,12pt,here,oneside,final]{article}
\usepackage{graphicx}
\usepackage{bm}
\usepackage{amsmath}
\usepackage{amsfonts}
\usepackage{amssymb}

\newcommand{\Slash}[1]{{\ooalign{\hfil/\hfil\crcr$#1$}}}

\begin{document}

\baselineskip 0.9cm

\begin{center}

{\LARGE Chirality imbalance and chiral magnetic effect under a parallel electromagnetic field}

\vspace{1cm}

{\large Hayato~Aoi\footnote{1218701@ed.tus.ac.jp} and Katsuhiko~Suzuki\footnote{katsu\_suzuki@rs.tus.ac.jp}}

\vspace{0.5cm}

{\it Department of Physics, Tokyo University of Science, \\Kagurazaka 1-3, Shinjuku, Tokyo 162-8601, Japan}

\vspace{2cm}

{\bf Abstract}






\end{center}

\date{\today}

We study the time evolution of the chirality
imbalance $n_5$ and the chiral magnetic effect (CME) under the  external parallel 
electromagnetic fields without assuming the artificial chiral
asymmetric source.   
We adopt the time-dependent Sauter-type
electric and constant magnetic field, and obtain analytical
solutions of the Dirac equation for a massive fermion.  We use the
point-split regularization to calculate the vacuum contribution in the
gauge invariant way.  As a result, we find that $n_5$ and CME current increase
substantially as the electric field increases, and stay finite after
the electric field is switched off.   The chirality imbalance and CME current are shown to
consist of a dominant contribution, which is essentially proportional to
relativistic velocity, and a small oscillating part.  We find a
simple analytical relation between $n_5$ and the fermion pair-production rate from the vacuum.   
We also discuss dynamical origin of the chirality imbalance in detail.  




\newpage

\baselineskip 0.7cm

\section{Introduction}
Recently, roles of the chiral anomaly have attracted considerable theoretical and  experimental interests 
in various subjects of physics.  
The chiral (Adler-Bell-Jackiw) anomaly is  
violation of the (partial) axial-vector current conservation due to quantum effects\cite{ABJ,anomaly},  
and causes the CP-violating processes observed experimentally.  
For the last decade, macroscopic manifestations of the chiral anomaly are discussed in the
context of hydrodynamic and transport 
phenomena in  systems with chiral fermions,  e.g.~the quark-gluon plasma or the 
Dirac$/$Weyl semi-metals\cite{transport1,ReviewT,transport2}.    
One of the important effects induced by the anomaly is  chiral magnetic effect (CME), which is the generation of 
''non-dissipative'' electric current along the direction of the magnetic field\cite{CME1,CME2,CME3};  
\begin{equation}
\displaystyle{{\bm J} =\frac{\mu_ 5}{2 \pi^2}  {\bm B}}
\label{CME}
\end{equation}
where $\mu_5$ is the chiral chemical potential.  
The chiral chemical  potential characterizes an asymmetry of the chirality of the system, 
and is conjugate to the chirality imbalance of the fermions, $n_5$, which is a difference of 
right-handed and left-handed fermion number densities, 
$n_5 \equiv n_R  - n_L  \equiv \langle \bar \psi \gamma_0 \gamma_5 \psi \rangle$.    

In the quark-gluon plasma produced in the heavy ion collisions,   
the interaction with the non-trivial gluonic field would change quark chiralities 
and thus produce   
the chirality imbalance between right- and left-handed quarks\cite{ReviewT}.  
With the strong magnetic field, $eB\sim m_\pi^2$,  created by the heavy ion-collision, CME may produce 
asymmetry of the 
charged particle distributions which can be measured experimentally\cite{ReviewE}.  
On the other hand, CME is an important topic in the condensed matter system\cite{transport2}, 
where the massless Dirac mode 
has been realized in the Dirac/Weyl semimetals\cite{Wmetal,ZeTr}.  
Experimental result for observing CME in such a system is reported in ref.~\cite{ZeTr}.

For various applications in QCD/condensed matter, the existence of the chirality imbalance $n_5$ and 
the chiral chemical potential $\mu_5$ is 
{\em a priori} assumed to study specific transport phenomena.  
However, appearance of the initial chirality imbalance is still under debate.  
For example, in the quark-gluon plasma,  metastable local CP-violating domains may be generated by 
transitions of the  non-perturbative gluonic configurations\cite{CME2,gluon,latticeCME}.     
In ref.~\cite{ZeTr} for the semimetal system with the electromagnetic field, $\mu_5$ is estimated as
$\displaystyle{\mu_5 = \hbar v_F \left(\frac{3e^2}{4}{\bm E}\cdot {\bm B} \tau \right)} \:,$
where $\tau$ is the relaxation time of the chirality imbalance.   
In our opinion, it is important to calculate the chirality imbalance and the chiral magnetic effect within the field theoretical method
without introducing additional assumptions.  

On the other hand, it is also necessary  to clarify use of  eq.~(\ref{CME})  in equilibrium.  
Although the CME formula eq.~(\ref{CME}) is used for various applications, 
it has to be interpreted with care.   
It is pointed out that such a current is forbidden in equilibrium system\cite{Bloch1,Bloch2}.   
There are also some cautions from theoretical calculations\cite{claims1,claims2}.   
It seems that the introduction of the chiral chemical potential implicitly assumes a system out of equilibrium\cite{CM-Schwinger1}.  
In order to clarify this issue, it is crucial to calculate time-evolution of the chirality imbalance and the 
CME current 
within a specific model and compare their characteristic time scales.

In addition, the CME current for a massive fermion is to be studied carefully.   
It is well-known that the anomaly relation receives a contribution from the mass-dependent term:
\begin{align}
\partial_\mu \int d^3 x \langle\bar{\psi}\gamma^\mu \gamma^5\psi\rangle 
=2im \int d^3 x \langle\bar{\psi} \gamma^5\psi\rangle + 
\frac{2 \alpha }{\pi} \int d^3 x {\bm E} \cdot {\bm B}  \, ,
\label{anomly}
\end{align}  
with $\alpha = e^2/4\pi$ being the fine structure constant.   To estimate the contribution from the mass, one should 
calculate a vacuum expectation value of the pseudoscalar density, which is also 
time-dependent.   
In particular, it is of interest  to understand a relation between CME and the spontaneous breakdown of 
chiral symmetry.   In ref.~\cite{Bchiral}, the CME current is suppressed  in the insulator phase, which 
may correspond to the chiral symmetry breaking phase.

For these purposes, we study 
time evolution of the chirality
imbalance $n_5$ and the chiral magnetic effect in the vacuum 
under the electromagnetic field solving the Dirac equation analytically without initial chiral chemical potential\cite{AoiSuzuki1}.  
We consider the vacuum state (zero temperature and zero fermion chemical potential) with external parallel 
electromagnetic fields, which provide the chirality imbalance of the fermion number density due to the chiral anomaly.  
We adopt the time-dependent half-pulse electric field (Sauter-type), and constant magnetic field in order to 
solve the Dirac equation for a massive fermion analytically.  
To calculate the infinite vacuum contribution in the gauge invariant way, we use the
point-split regularization\cite{regularization,Warringa:2012bq} and calculate vacuum expectation values of the 
various bilinear fermion operators including $n_5$ and CME.   
In addition,    
we expect production of  fermion-antifermion pairs from the vacuum under the electric field by 
the Schwinger mechanism\cite{Schwinger,SchBog}.  
We systematically study relations  
between  $n_5$, CME current and the pair-creation rate using the 
Bogoliubov transformation.   
Our results are to be compared with the previous works obtained with the Schwinger 
mechanism with constant electromagnetic field\cite{CM-Schwinger1}, 
the Wigner function method with collinear electromagnetic fields\cite{Wigner1}, 
the Wigner function method with the chiral chemical potential\cite{Wigner2}, 
cylindrical Dirac equation with the chiral chemical potential\cite{Fukushima2}.

This paper is organized as follows.
In Sec.~\ref{sec:dirac-equat-squar}, we 
show  analytical solutions of the Dirac equation with the 
parallel electromagnetic fields.   Using them we perform the canonical quantization and define the vacuum state at 
$t\to {-\infty}$ in Sec.~\ref{sec:quantization}.  
We introduce the point-split regularization in Sec.~\ref{section:Regularization and Chiral Anomaly} to calculate vacuum expectation values of the fermion operators in gauge invariant way.  
We present our numerical results for the time evolution of the vacuum expectation values of the chirality imbalance and CME 
in Sec.~\ref{section:Time evolution of  the VEVs of the currents}.   In Sec.~\ref{Relation to the fermion pair-prudcution}   
we also discuss relations between the chirality imbalance and the fermion pair-production rate, and give a simple formula for CME 
current.    Using them, we show how the chirality imbalance is dynamically generated in this model.  
Finally, Sec.~\ref{label:Summary and Disscussions} is devoted to the summary and discussion.

\section{Solution of Dirac eq. under solvable external EM fields}
\label{sec:dirac-equat-squar}
\subsection{Dirac equation with electromagnetic fields}

We need an analytical solution of the Dirac equation under the constant magnetic and the time-dependent 
electric field, which plays a key role in our work.  
The Dirac equation for a fermion field $\psi(x)$ 
with the mass $m$ under an external electromagnetic potential, $A_{\mu}(x)$, is given by 
\begin{align}
  &[i\Slash{D}-m]\psi(x)=0
  \label{eq:dirac-eq-co-deriv}
\end{align}
where we introduce the  covariant derivative $ D_{\mu}=\partial_{{\mu}}+ieA_{\mu}(x)$.  
Squared Dirac equation is given by
\begin{align}
  & \Slash{D}^2\Phi(x)=-m^2\Phi(x) \label{eq:dirac-eq-z-0}
\end{align}  
Hereafter, we concentrate on finding solutions of the squared Dirac equation $\Phi(x)$, from which 
we can obtain $\psi(x)$  by a suitable projection,
\begin{align}
  \psi(x)=(i\Slash{D}+m)\Phi(x) \; .
    \label{eq:dirac-eq-ansatz}
\end{align}


We consider specific forms of the  external electromagnetic field in this work to obtain analytical solutions;  
\begin{align}
\boldsymbol{B}&=(0, 0, -B), \\
\boldsymbol{E}&=(0, 0, E/\cosh^2(t/\tau)) \; , 
\label{Sauter}
\end{align}
where parameters $B$ and $E$ are non-zero real constants, and $\tau>0$.  
The magnetic field is time-independent and uniform along the $z$-direction. On the other hand,  the electric field is spatially homogeneous 
but time-dependent with a pulse structure, which is known as Sauter-type electric field\cite{Sauter}.    
The corresponding electromagnetic potential is 
\begin{align}
	A_{\mu}=(0, 0, Bx, E\tau (\tanh(t/\tau)+1)) \label{eq:em-potential} \; .  
\end{align}
Note that the vector potential is finite even at $t \to \pm \infty$.   
If we adopted the constant electric field\cite{SchBog,Warringa:2012bq}, the vector potential would diverge at $t \to \pm \infty$, $A_3 = E\, t$.    

To understand roles of the electromagnetic field,  it is convenient to introduce the so-called "magnetic helicity" density\cite{MagneticHelicity} 
as 
\begin{align}
 h(t) \equiv \frac{1}{V}\int d^3 x \boldsymbol{A}\cdot \boldsymbol{B} \; .
\label{MH}
\end{align}
Although the magnetic helicity is not gauge invariant in general, 
it is useful when we discuss the topological structure of the gauge field.  
With eq.~(\ref{eq:em-potential}), the magnetic helicity density in our case is calculated as
\begin{align}
  h(t) =-BE\tau(\tanh(t/ \tau)+1)\; .  
\end{align}
At the initial state, $t\to -\infty$, both electric field and the magnetic helicity $h(t)$ are zero,  
and they increase as  $t$ increases.    
In the final state, $t \to\infty$, ${\bm E}$ vanishes rapidly, while the magnetic helicity $h(t)$ 
is kept finite.   
This peculiar behavior of the magnetic helicity is due to the Sauter electric field, and appropriate to discuss the production of 
the chirality imbalance, as we will show later.  



With the chiral representation for the gamma matrices, 
the Dirac operator $\Slash{D}$ and  its squared form  $\Slash{D}^2$ are given by
\begin{align*}
  i\Slash{D}
  &=
	\begin{pmatrix}
          0 & 0 & -\hat{c}_{-} & i\hat{a} \\
          0 & 0 & -i\hat{a}^{\dagger}  &  \hat{c}_{+} \\
          \hat{c}_{+} & -i\hat{a} & 0 & 0 \\
          i\hat{a}^{\dagger}  & -\hat{c}_{-} & 0 & 0 
        \end{pmatrix} \\
  \Slash{D}^2 
  &= \begin{pmatrix}
          \hat{c}_{-}\hat{c}_{+} + \hat{a}\hat{a}^{\dagger}& 0 & 0 & 0 \\
          0 & \hat{c}_{+}\hat{c}_{-} + \hat{a}^{\dagger}\hat{a} & 0  &  0 \\
           0 & 0 & \hat{c}_{+}\hat{c}_{-} +\hat{a}\hat{a}^{\dagger} & 0 \\
          0  & 0 & 0 &  \hat{c}_{-}\hat{c}_{+} +\hat{a}^{\dagger}\hat{a}
      \end{pmatrix} 
\end{align*}
where we have defined the following operators,
\begin{align*}
	\hat{c}_{+} &=(-i\partial_z+eE\tau(\tanh(t/\tau)+1))+i\partial_t\\
	\hat{c}_{-} &=(-i\partial_z+eE\tau(\tanh(t/\tau)+1))-i\partial_t\\
	\hat{a} &=(-i\partial_y+eBx)+\partial_x\\
	\hat{a}^{\dagger} &=(-i\partial_y+eBx)-\partial_x \; .
\end{align*}
Because 
$\Slash{D}^2$ commute both $\partial_{y}$ and $\partial_{z}$, 
the solution of the squared Dirac equation, $\Phi$, 
can be  written as a separable form, $\Phi(t, \boldsymbol{x})=\exp(ip_y y + ip_z z)\, \phi(t,x)$, 
with momenta of  $y$ and $z$ directions being constants.  
For  $\Phi(t, {\bm x})$, we explicitly introduce the four component form as
\begin{align}
  \phi(t,x)=
  \begin{pmatrix}
    \phi_1(t,x) \\
    \phi_2(t,x) \\
    \phi_3(t,x) \\
    \phi_4(t,x) 
  \end{pmatrix}\; .
\end{align}
We then obtain a set of equations for $\phi_i(t,x)~~(i=1,2,3,4)$ as follows,
\begin{align}
&[\hat{c}_{-}\hat{c}_{+}+\hat{a}\hat{a}^{\dagger}+m^2]\phi_1(t,x)=0\\
&[\hat{c}_{+}\hat{c}_{-}+\hat{a}^{\dagger}\hat{a}+m^2]\phi_2(t,x)=0\\
&[\hat{c}_{+}\hat{c}_{-}+\hat{a}\hat{a}^{\dagger}+m^2]\phi_3(t,x)=0\\
&[\hat{c}_{-}\hat{c}_{+}+\hat{a}^{\dagger}\hat{a}+m^2]\phi_4(t,x)=0 \; . 
\end{align}
Note that the operators $\hat{c}_{+}, \hat{c}_{-}$ include only $t$ and $\partial_t$ variables, whereas 
$\hat{a}, \hat{a}^{\dagger}$ contain only $x$ and $\partial_x$.  Hence, 
these equations can be solved as 
\begin{align}
	\Phi({\bm x})=
	\exp(ip_y y + ip_z z)
	\begin{pmatrix}
	f_1(t)g_1(x) \\
	f_2(t)g_2(x) \\
	f_2(t)g_1(x) \\
	f_1(t)g_2(x) 
\end{pmatrix} 
\label{eq_tot}
\end{align}
with eigenfunctions, $f_i(t), g_i(x) \; (i=1, 2)$, which satisfy the following eigenvalue equations, 
\begin{align}
&\hat{a} \, \hat{a}^{\dagger} g_1(x)=\kappa g_1(x) \label{eq-g1}\\
&\hat{a}^{\dagger} \, \hat{a} g_2(x)=\kappa g_2(x) \label{eq-g2}\\
&\hat{c}_{-}\, \hat{c}_{+}f_1(t)=-(\kappa+m^2)f_1(t) \label{eq-f1}\\
&\hat{c}_{+}\, \hat{c}_{-}f_2(t)=-(\kappa+m^2)f_2(t) \label{eq-f2} \; .
\end{align}
We note that 
the eigenvalue $\kappa$ is real and positive-semidefinite because 
the operators $\hat{a}^\dagger \hat{a}, \hat{a}\hat{a}^\dagger$ are Hermittian.  

\subsection{solutions for the  $x$-dependent part}
\label{sec:x-part}

The eigenfunction and the eignevalue of equations (\ref{eq-g1}) and (\ref{eq-g2}) are easily obtained with the 
standard technique for the  harmonic oscillator.  

We find a solution with the normalized Hermite polynomial $\mathrm{H}_n$ $(n=0, 1, 2, \cdots)$
\begin{align}
	g_1(x)&= g_{n-1, p_y}(x)\\
	g_2(x)&=  g_{n, p_y}(x)
\end{align}
with eigenvalues $\kappa=2eBn$ and  eigenfunctions $g_{n,p_y}(x)$; 
\begin{align}
  g_{n,p_y}(x) &=  \frac{1}{\sqrt{2^n n!}}(\frac{|eB|}{\pi})^{1/4}H_n(\eta)\exp(-\eta^2/2) \label{eq:normalized-Hermite-func}\\  
\eta  &\equiv \frac{1}{\sqrt{|eB|}}(p_y+|eB|x)  \; \, .\nonumber                
\end{align}
where $n$ denotes the Landau level.  
When $n=0$, the normalizable solution of $g_{1}(x)$ does not exist, thus we define $g_{-1,p_y}(\tilde{x})=0$.

The eigenfunctions satisfy the orthonormal condition as 
\begin{align}
  &\int d x \, g_{n,p_y}(x)g_{m,p_y}(x)=\delta_{mn} \;  .
\end{align}
Moreover, the completeness identity also holds
\begin{align}
  \sum_{n=0}^{\infty}g_{n,p_y}(x)g_{n,p_y}(x')=\delta(x-x') \; .
\end{align}
Additionally, integration over $p_y$ also gives a relation
\begin{align}
  &\int d p_y \, g_{n,p_y}(x)g_{m,p_y}(x)=|eB|\delta_{mn} 
\end{align}
which guarantees the orthogonal condition for $y$ in eq.~(\ref{eq_tot}).


\subsection{solutions for the $t$-dependent part}
\label{sec:t-part}

Next, we will solve equations for the time-dependent part.  
The operators $\hat{c}_{-}\hat{c}_{+}, \hat{c}_{+}\hat{c}_{-}$ in eq.~(\ref{eq-f1}) and  eq.~(\ref{eq-f2}) are written explicitly as 
\begin{align}
  &\hat{c}_{+}\hat{c}_{-} = \partial_t^2+(p_z+eE\tau (\tanh(t/ \tau)+1))^2+ie\frac{E}{\cosh^2(t/\tau)} \label{eq-c1}\\ 
  &\hat{c}_{-}\hat{c}_{+} =\partial_t^2+(p_z+eE\tau (\tanh(t/ \tau)+1))^2-ie\frac{E}{\cosh^2(t/\tau)}\label{eq-c2} \; \; .
\end{align}
which reduce to the hypergeometric differential equation for $f_1(x)$.  
We obtain the eigenfunctions for  $f_1(x)$ in (\ref{eq-f1}) and $f_2(x)$ in (\ref{eq-f2}) as follows
\begin{align}
  & \tilde{\phi}_{n,p_z}^{(+)}(t) \equiv \sqrt{\frac{\omega(0)+p_z}{2\omega(0)}}u^{-\frac{i\tau\omega(0)}{2}}(1-u)^{\frac{i\tau\omega(1)}{2}}F\left({a,b \atop c};u(t)\right)\label{phi_p}\\
  &\tilde{\phi}_{n,p_z}^{(-)}(t) \equiv \sqrt{\frac{\omega(0)-p_z}{2\omega(0)}}u^{\frac{i\tau\omega(0)}{2}}(1-u)^{-\frac{i\tau\omega(1)}{2}}\nonumber \\
& \hspace {2.5cm}\times F\left({1-a,1-b \atop 2-c};u(t)\right)\label{phi_m}
\end{align}
where $F\left({a,b \atop c};u\right)$ are Gauss's hypergeometric function.  The parameters $a,b,c$ are given by 
\begin{subequations}
\begin{align}
  &a=1-\frac{i\tau\omega_{n,p_z}(0)}{2}+\frac{i\tau\omega_{n,p_z}(1)}{2}+ieE\tau^2\\
  &b=-\frac{i\tau\omega_{n,p_z}(0)}{2}+\frac{i\tau\omega_{n,p_z}(1)}{2}-ieE\tau^2 \\
  &c=1-i\tau\omega_{n,p_z}(0) \; ,
\end{align}
where
\end{subequations}
\begin{align}
  \omega_{n,p_z}^2(u)&=(p_z+2eE\tau u)^2+2|eB|n+m^2 \label{omega_def} \\
  u(t)&=\frac{1}{2}(\tanh(t/ \tau)+1)    \; .
\end{align}
We find a simple relation, 
\begin{align}
  |\tilde{\phi}_{n,p_z}^{(+)}(t)|^2+|\tilde{\phi}_{n,p_z}^{(-)}(t)|^2=1, 
\end{align}
which holds independent of  $t$, and is useful for further  calculations.    


\subsection{classical solutions of Dirac equation}
\label{sec:class-solut-dirac-1}

We then obtain  solutions of the squared Dirac equation, $\Phi_{n,p_y,p_z}(x)$, as follows:
\begin{align}
&  \Phi_{n,p_y,p_z} =
    \exp(ip_y y + ip_z z) \nonumber \\
&\hspace{0.5cm} 
\times     \begin{pmatrix}
    g_{n-1,p_y}(x)\{ N^{(+)}_{1}\tilde{\phi}_{n, p_z}^{*(-)}(t) + N^{(-)}_{1}\tilde{\phi}_{n, p_z}^{*(+)}(t)\} \\
    g_{n,p_y}(x)\{ N^{(-)}_{2}\tilde{\phi}_{n, p_z}^{(-)}(t) + N^{(+)}_{2}\tilde{\phi}_{n, p_z}^{(+)}(t)\} \\
    g_{n-1,p_y}(x)\{ N^{(-)}_{3}\tilde{\phi}_{n, p_z}^{(-)}(t) + N^{(+)}_{3}\tilde{\phi}_{n, p_z}^{(+)}(t)\} \\
    g_{n,p_y}(x)\{N^{(+)}_{4} \tilde{\phi}_{n, p_z}^{*(-)}(t) +N^{(-)}_{4} \tilde{\phi}_{n, p_z}^{*(+)}(t)\}
  \end{pmatrix}
 \label{sd_sol0}
\end{align}
where $N^{(\pm)}_{i}$ are normalization constants.  
To construct the solutions of the Dirac equation, we properly choose solutions in 
eq.~(\ref{sd_sol0}), and extract the right/left-handed solutions by performing  the suitable projection.  
Here, we choose 4-independent solutions proportional to $N^{(\pm)}_{1}, N^{(\pm)}_{4}$ in eq.~(\ref{sd_sol0}), 
which satisfy the orthogonal and completeness relations, as we will show later.  

First, we obtain the "right-handed" solutions, operating $i\gamma^\mu D_\mu+m$ to the 1st row in eq.~(\ref{sd_sol0}).
\begin{align}
  \psi_{\boldsymbol{p}}^{(+,\tilde{R})}
  &=N^{(+)}_{1}\exp(ip_y y + ip_z z)\nonumber \\
&\times  
    \begin{pmatrix}
      \cos\theta_n \cdot g_{n-1,p_y}(x)\cdot \tilde{\phi}_{n, p_z}^{*(-)}(t)\\
      0 \\
      g_{n-1,p_y}(x)\cdot  \tilde{\phi}_{n, p_z}^{(+)}(t)\\
      i \sin\theta_n \cdot  g_{n,p_y}(x)  \cdot \tilde{\phi}_{n, p_z}^{*(-)}(t)
    \end{pmatrix} \label{sol_right1}
  \\
  \psi_{\boldsymbol{p}}^{(-,\tilde{R})}
  &=N^{(-)}_{1}\exp(ip_y y + ip_z z)\nonumber \\
& \times    \begin{pmatrix}
      \cos\theta_n \cdot  g_{n-1,p_y}(x) \cdot \tilde{\phi}_{n, p_z}^{*(+)}(t)\\
      0 \\
      -g_{n-1,p_y}(x)\cdot  \tilde{\phi}_{n, p_z}^{(-)}(t)\\
      i \sin\theta_n \cdot g_{n,p_y}(x)  \cdot \tilde{\phi}_{n, p_z}^{*(+)}(t)
    \end{pmatrix} \label{sol_right2}
\end{align}
where $\theta_n$ is defined by
\begin{align}
  \theta_{n}=\arctan(\frac{\sqrt{2|eB|n}}{m})  \; .
 \label{sol_right3}
\end{align}
Hereafter, we use the shorthand notation $\boldsymbol{p}=(n, p_y, p_z)$ for simplicity.

Similarly, we obtain the "left-handed" solutions  operating  $i\gamma^\mu D_\mu+m$ to the 4th row in  eq.~(\ref{sd_sol0}).   
\begin{align}
  \psi_{\boldsymbol{p}}^{(+,\tilde{L})}
  &=N^{(+)}_{4}\exp(ip_y y + ip_z z)\nonumber \\
& \times     \begin{pmatrix}
      i \sin\theta_n \cdot g_{n-1,p_y}(x)  \cdot \tilde{\phi}_{n, p_z}^{*(-)}(t)\\
      g_{n,p_y}(x)\cdot  \tilde{\phi}_{n, p_z}^{(+)}(t) \\
      0 \\
      \cos\theta_n \cdot g_{n,p_y}(x) \cdot  \tilde{\phi}_{n, p_z}^{*(-)}(t)
    \end{pmatrix} \label{sol_left1}
  \\
  \psi_{\boldsymbol{p}}^{(-,\tilde{L})}
  &=N^{(-)}_{4}\exp(ip_y y + ip_z z) \nonumber \\
&\times    \begin{pmatrix}
      i \sin\theta_n \times g_{n-1,p_y}(x)  \cdot \tilde{\phi}_{n, p_z}^{*(+)}(t)\\
      -g_{n,p_y}(x)\cdot  \tilde{\phi}_{n, p_z}^{(-)}(t) \\
      0 \\
      \cos\theta_n \cdot g_{n,p_y}(x) \cdot \tilde{\phi}_{n, p_z}^{*(+)}(t)
    \end{pmatrix} \label{sol_left2}
\end{align}
In massless limit, $m \to 0$, the solutions of eq.~(\ref{sol_right1})  
are exact eigenspinors of the chirality operator $\gamma^5$ with the eigenvalue $+1$, while 
eq.~(\ref{sol_left1}) is the chirality eigenstate with the eigenvalue $-1$.    
Note that $\psi_{0, p_y, p_z}^{(\pm,\tilde{R})}=0$ because of $g_{-1, p_y}=0$ and $\sin\theta_0=0$.

These solutions of the  Dirac equation form the complete orthonormal basis.  
By choosing the normalization constants, $N^{(+)}_{1}=N^{(-)}_{1}=N^{(+)}_{4}=N^{(-)}_{4}=1$, 
the orthonormal relations are given by
\begin{align*}
  &\int d^3 x [\psi_{\boldsymbol{p'}}^{(u', s')}(x)]^{\dagger}[\psi_{\boldsymbol{p}}^{(u, s)}(x)\nonumber \\
& \hspace{1cm}   =(2\pi)^2\delta_{uu'}\delta_{ss'}\delta_{nn'}\delta(p_y-p_y')\delta(p_z-p_z') ; 
\end{align*}
which holds except for  $\psi_{0, p_y, p_z}^{(\pm,\tilde{R})}$.  
Moreover, one can show the completeness relation for eqs.~(\ref{sol_right1},\ref{sol_right2},\ref{sol_left1},\ref{sol_left2})  
\begin{align*}
  &\sum_{\boldsymbol{p}} \sum_{s=\tilde{R}, \tilde{L}}\sum_{u=\pm}
    [\psi_{\boldsymbol{p}}^{(u,s)}(t,\boldsymbol{x})]_{\alpha}[\psi_{\boldsymbol{p}}^{\dagger(u, s)}(t,\boldsymbol{x}')]_{\beta}
\nonumber \\
& \hspace{1cm} =(2\pi)^2\delta_{\alpha\beta}\delta^{(3)}(\boldsymbol{x}-\boldsymbol{x}') \; ,
\end{align*}
which guarantees the validity of  our construction from eq.~(\ref{sd_sol0}).

\section{Quantization and vacuum expectation values of currents}
\label{sec:quantization}

To construct the quantum field theory with the external EM field, we first introduce
the fermionic field operators from eqs.~(\ref{sol_right1}), (\ref{sol_right2}), (\ref{sol_left1}), (\ref{sol_left2}); 
\begin{align}
  \hat{\psi}(x)
  &= \sum_{n=0}^{\infty}\int \frac{d p_y}{\sqrt{2\pi}} \int \frac{d p_z}{\sqrt{2\pi}} \nonumber \\
&\times \sum_{s=\tilde{R}, \tilde{L}}
    (\hat{b}_{s,\boldsymbol{p}}^{}\psi_{s,\boldsymbol{p}}^{(+)}(x)+\hat{d}_{s,-\boldsymbol{p}}^{\dagger }\psi_{s,\boldsymbol{p}}^{(-)}(x))
\end{align}
where $\hat{b}_{s,p}^{\dagger}, \hat{d}_{s,p}^{\dagger}$($\hat{b}_{s,p}, \hat{d}_{s,p}$) are interpreted as creation (annihilation) operators of the 
particles and anti-particles.  
These operators obey the anti-commutation relations, 
\begin{align*}
  &\{\hat{b}_{s,p}, \hat{b}_{s',p'}^{\dagger}\}
    =\{\hat{d}_{s,p}, \hat{d}_{s',p'}^{\dagger}\}=\delta_{ss'}\delta_{nn'}\delta(p_y-p_y')\delta(p_z-p_z')
\end{align*}
which is equivalent to the anti-commutation relations for the field operators, 
\begin{align*}
  \{\hat{\psi}_{\alpha}(t,\boldsymbol{x}),\hat{\psi}^{\dagger}_{\beta}(t,\boldsymbol{x}')\}=\delta^{(3)}(\boldsymbol{x}-\boldsymbol{x}')
\delta_{\alpha\beta} \; .
\end{align*}

In order to describe the fermion field under the  time-dependent EM field, 
we adopt the Heisenberg picture in the following calculations, and define the 
vacuum state $|0\rangle$ at $t \to -\infty$, 
\begin{align}
  \hat{b}_{s,\boldsymbol{p}}|0\rangle
  = 0, ~\hat{d}_{s,\boldsymbol{p}}|0\rangle=0~(\text{for all }s,\boldsymbol{p}),~~\langle 0|0\rangle=1 \; .
\end{align}
We obtain asymptotic behavior of the eigenfunctions 
 $\tilde{\phi}_{n,p_z}^{(\pm)}(t)$ at $t \to -\infty$ as  
\begin{subequations}
 \begin{align}
  &\tilde{\phi}_{n,p_z}^{(+)}(t) \propto \exp(-i\omega_{n,p_z}(0)t) ~~~(t\to -\infty)
    \label{eq:f-in-plus-asymptotic-free-field}\\
  &\tilde{\phi}_{n,p_z}^{(-)}(t) \propto \exp(+i\omega_{n,p_z}(0)t) ~~~(t\to -\infty)
    \label{eq:f-in-minus-asymptotic-free-field}
\end{align} 
\end{subequations}
Apparently, the eigenfunction $\tilde{\phi}_{n,p_z}^{(+)}(t)$ ($\tilde{\phi}_{n,p_z}^{(-)}(t)$) at $t \to -\infty$ coincides 
a positive (negative) energy solution of the free Dirac fermion.

By using the quantized fields, the classical current, $j(\Gamma;x)=\bar{\psi}(x)\Gamma\psi(x)$, is  replaced by the current operator 
\begin{align*}
  \hat{j}({\Gamma};x)
  &=\frac{1}{2}[\hat{\bar{\psi}}(x), \Gamma\hat{\psi}(x)]\\
  &=\frac{1}{2}[\hat{\bar{\psi}}_{\alpha}(x)\Gamma_{\alpha\beta}\hat{\psi}_{\beta}(x)-\Gamma_{\alpha\beta}\hat{\psi}_{\beta}(x)\hat{\bar{\psi}}_{\alpha}(x)]
\end{align*}
where $\Gamma$ are products of $\gamma$ matrices, i.e.~
$\Gamma = (1, i \gamma_5, \gamma_\mu, \gamma_5 \gamma_\mu) $.    
We can calculate the vacuum expectation value (VEV) of the corresponding current as follows
\begin{subequations}
\begin{align}
  \langle \bar{\psi}(x)\Gamma\psi(x)\rangle
  &= \langle 0 |\hat{j}({\Gamma};x)|0\rangle \label{eq:current-expectation-1}\\
  & =\sum_{n=0}^{\infty}\int \frac{d p_y}{2\pi} \int \frac{d p_z}{2\pi} S_{\boldsymbol{p}}(x;\Gamma)\label{eq:current-expectation-2} 
\end{align}  
\end{subequations}
where we define $S_{\boldsymbol{p}}(x;\Gamma)$ as
\begin{align}  
  S_{\boldsymbol{p}}(x;\Gamma)& \equiv \frac{1}{2}\sum_{s=\tilde{R},\tilde{L}}[\bar{\psi}_{\boldsymbol{p}}^{(-,s)}(x)\Gamma\psi_{\boldsymbol{p}}^{(-,s)}(x)   \nonumber \\
&\hspace{1cm} - \bar{\psi}_{\boldsymbol{p}}^{(+,s)}(x)\Gamma\psi_{\boldsymbol{p}}^{(+, s)}(x)] \label{eq:a}
\end{align}  

Using eqs.~(\ref{sol_right1}),(\ref{sol_right2}),(\ref{sol_left1}),(\ref{sol_left2}) we find  $S_{\boldsymbol{p}}(x;\Gamma)$ 
for  various $\Gamma$ as,
\begin{align}  
  &S_{\boldsymbol{p}}(x;\gamma^0\gamma^5)
    = [g_{n-1, p_y}^2-g_{n, p_y}^2][|\tilde{\phi}_{n, p_z}^{(+)}|^2-|\tilde{\phi}_{n, p_z}^{(-)}|^2]
\label{n5}
\\
  &S_{\boldsymbol{p}}(x;\gamma^3)
    = [g_{n-1, p_y}^2+g_{n, p_y}^2][|\tilde{\phi}_{n, p_z}^{(+)}|^2-|\tilde{\phi}_{n, p_z}^{(-)}|^2]
\label{v3}
\\
  &S_{\boldsymbol{p}}(x;i\gamma^5)
    = 2[g_{n-1, p_y}^2-g_{n, p_y}^2]\cos\theta_n\, \mbox{Im}[\tilde{\phi}_{n, p_z}^{(+)} \phi_{n, p_z}^{(-)}]
\label{g5}\\
  &S_{\boldsymbol{p}}(x;\gamma^0\gamma^3)
    = 2[g_{n-1, p_y}^2+g_{n, p_y}^2]\cos\theta_n\, \mbox{Im}[\tilde{\phi}_{n, p_z}^{(+)} \phi_{n, p_z}^{(-)}] 
\label{av3}
\\   
  &S_{\boldsymbol{p}}(x;i\gamma^1\gamma^2)
    = 2[g_{n-1, p_y}^2-g_{n, p_y}^2] \cos\theta_n\, \mbox{Re}[\tilde{\phi}_{n, p_z}^{(+)} \phi_{n, p_z}^{(-)}] 
\label{g12}
\\
  &S_{\boldsymbol{p}}(x;1)
    = 2[g_{n-1, p_y}^2+g_{n, p_y}^2]\cos\theta_n\mbox{Re}[\tilde{\phi}_{n, p_z}^{(+)} \phi_{n, p_z}^{(-)}]  
\label{sca1} \; , 
\end{align} 
For further calculations, we shall integrate the right hand side over $p_y$ with paying attention to $g_{-1,p_y}=0$, namely, 
\begin{align}
  \int d p_y[g_{n-1, p_y}^2(x)-g_{n, p_y}^2(x)]&=-|eB|\delta_{n0}\\
  \int d p_y[g_{n-1, p_y}^2(x)+g_{n, p_y}^2(x)]&=|eB|\alpha_n
\end{align}
where $\alpha_n$ are defined by
\begin{align}
  \alpha_n =
  \begin{cases}
     1  & \text{if~}n=0 \\
     2  & \text{if~}n =1,2,3,\cdots \; .
    \end{cases}
\end{align}


\section{Regularized VEV and  Chiral Anomaly}
\label{section:Regularization and Chiral Anomaly}

\subsection{Regularization and and VEVs of currents}

The VEVs of the current derived in the previous section diverge when we integrate over  $p_z$, thus 
 we need some sort of  the regularization.    
Because we could obtain these VEVs as a result of the subtle cancellation of the
divergent integrals, use of the gauge invariant regularization is certainly  important.   
Here, we use the point-split regularization\cite{regularization,Warringa:2012bq}, which is known as the gauge invariant regularization scheme.

The regularization method in the $p_z$ integral essentially  introduces the non-locality in the $z$ space.  
We replace the local current operator, $\bar{\psi}(x)\Gamma{\psi}(x)$, by the integral of the non-local current
as follows
\begin{align}
  \bar{\psi}(z)\Gamma{\psi}(z)
  &=\int dz'\bar{\psi}(z')\delta(z-z')\Gamma{\psi}(z)\\
  &=\lim_{\varepsilon \to 0}  \int dz'\bar{\psi}(z')h_{\varepsilon}(z-z')\Gamma{\psi}(z)
\end{align}
where 
\begin{align}
  h_{\varepsilon}(z-z')  \equiv  \frac{1}{2\sqrt{\pi\varepsilon}}\exp(-\frac{(z-z')^2}{4\varepsilon})
\end{align}
which is reduced to the delta function $\delta (z-z')$ as $\varepsilon \to 0$.

This non-locality clearly breaks the local gauge invariance of the matrix elements.   
To revover the gauge invariance, we insert the Wilson line into the non-local current
\begin{align}
  U(z',z)=\mbox{P} \, \exp[ie\int_{z'}^{z}d\tilde{x}^{\mu}A_{\mu}(\tilde{x})]
\end{align}
where the choice of the integral path $\mbox{P}$ is arbitrary.   
We may choose a straight line which connects $z'$ and $ z$ for the path function $U(z',z)$.  
\begin{align}
  \bar{\psi}(x)\Gamma{\psi}(x) \to \int_{-\infty}^{\infty}d z' \bar{\psi}(z')h_{\varepsilon}(z-z')U(z',z) \Gamma{\psi}(z) \; .
\end{align}
Carrying out the  $z'$ integration, we obtain the regularized VEV of the current, 
\begin{align}
  &\langle\bar{\psi}(x)\Gamma\psi(x)\rangle
    =\sum_{n=0}^{\infty}\int \frac{d p_y}{2\pi} \int \frac{d p_z}{2\pi}\nonumber \\
& \hspace{0.5cm}
\times \sum_{s=\tilde{R},\tilde{L}} \, \exp(-\varepsilon{[p_z+eA_z(t)]^2})\; S_{\boldsymbol{p}}(x;\Gamma)
\end{align}
The regularization factor, $\exp(-\varepsilon{[p_z+eA_z(t)]^2})$, is now inserted into the integrand for the VEVs.  
Because the parameter $\varepsilon$ has the dimension of  (mass)$^{-2}$,  
we introduce the cutoff parameter $\Lambda^2 \equiv 1/\varepsilon $ 
with the dimension of $\Lambda$ being (mass)$^1$.

We arrive at final expressions for the regularized VEVs of the currents:  .
\begin{align}
  &\langle\bar{\psi}\gamma^0\gamma^5\psi\rangle
    = \frac{eB}{2\pi}
    \int \frac{d p_z}{2\pi}  f_{\Lambda}(p_z)[|\tilde{\phi}_{0, p_z}^{(+)}(t)|^2-|\tilde{\phi}_{0, p_z}^{(-)}(t)|^2]
\label{n5_2}
\\
  &\langle\bar{\psi}i\gamma^1\gamma^2\psi\rangle
    =  2\frac{eB}{2\pi}
    \int \frac{d p_z}{2\pi} f_{\Lambda}(p_z)
    \cdot \mbox{Re}[\tilde{\phi}_{0, p_z}^{(+)}(t) \, \phi_{0, p_z}^{(-)}(t)]\label{g12r}\\
  &\langle\bar{\psi}i\gamma^5\psi\rangle
    =-2  \frac{eB}{2\pi}
    \int \frac{d p_z}{2\pi} f_{\Lambda}(p_z)
    \cdot\mbox{Im}[\tilde{\phi}_{0, p_z}^{(+)}(t) \,  \phi_{0, p_z}^{(-)}(t)] \label{PS_reg}\\
  &\langle\bar{\psi}\gamma^3\psi\rangle
    =  \frac{|eB|}{2\pi}  \int \frac{d p_z}{2\pi} \sum_{n=0}^{\infty} \alpha_n 
    f_{\Lambda}(p_z)[|\tilde{\phi}_{n, p_z}^{(+)}(t)|^2-|\tilde{\phi}_{n, p_z}^{(-)}(t)|^2] \label{CME_reg}\\
  &\langle\bar{\psi}\psi\rangle
    =  -2\frac{|eB|}{2\pi}  \int \frac{d p_z}{2\pi} \sum_{n=0}^{\infty} 
    \alpha_n f_{\Lambda}(p_z)\cos\theta_n\nonumber \\
& \hspace{3cm}  \times  \mbox{Re}[\tilde{\phi}_{n, p_z}^{(+)}(t) \, \tilde{\phi}_{n, p_z}^{(-)}(t)]\label{sca1r}\\
  &\langle\bar{\psi}i\gamma^0\gamma^3\psi\rangle
    =  2\frac{|eB| }{2\pi}  \int \frac{d p_z}{2\pi} \sum_{n=0}^{\infty} \alpha_n
    f_{\Lambda}(p_z) \cos\theta_n \nonumber \\
& \hspace{3cm} \times     \mbox{Im}[\tilde{\phi}_{n, p_z}^{(+)}(t) \, \tilde{\phi}_{n, p_z}^{(-)}(t)]
\end{align}
where $f_{\Lambda}(p_z)=\exp(-[p_z+eA_z(t)]^2 / \Lambda^2)$.     

It is clear that only the lowest Landau level (LLL), $n=0$, contributes to the 
chirality imbalance eq.~(\ref{n5_2}), psedoscalar density eq.~(\ref{PS_reg}), and the tensor density eq.~(\ref{g12r}).  
On the other hand, we need sum up contributions from all possible Landau levels for the vector current eq.~(\ref{CME_reg}) 
as well as the scalar density (\ref{sca1r}).   
This is different from calculations with the chiral chemical potential
\cite{Wigner2,Fukushima2},  
where the vector current is given by the contribution from only the lowest Landau level.    
In our case, when there is no electric field ($t=-\infty$), 
$|\tilde{\phi}_{n, p_z}^{(+)}|^2$ and $|\tilde{\phi}_{n, p_z}^{(-)}|^2$ in (\ref{CME_reg}) show the same momentum distribution 
for  each $n$, and a cancellation gives null vector current.  
After the electric field is turned on ($t\ge0$),  however, 
momentum distributions of $|\tilde{\phi}_{n, p_z}^{(+)}|^2$ and $|\tilde{\phi}_{n, p_z}^{(-)}|^2$ with the 
regularizatin function  become different in eqs.(\ref{phi_p},\ref{phi_m}), and 
the resulting vector current is finite for each $n$, although contributions from higher Landau levels are small.  
Our results are consistent with ref.~\cite{Warringa:2012bq}.

We also find that the VEVs for all other $\Gamma$s vanish.  
In particular, the spin expectation value of the $z$ component vanishes, 
$\langle \bar \psi \gamma_3\gamma_5\psi \rangle \sim \langle S_z\rangle = 0$.  
Thus, there is no magnetization of the vacuum due to the chiral anomaly.    
This is in contrast with the result for the tensor matrix element, 
$\langle \bar \psi \sigma_{12} \psi \rangle \neq 0$, in eq.~(\ref{g12r}), which is non-zero.  
This difference may come from roles of antifermions for these matrix elemets, i.e.~
$\langle \bar \psi \gamma_3\gamma_5\psi \rangle$ expresses a sum of fermion and antifermion contributions,
while the tensor matrix element describes their differences.


We are particularly interested in the $z$-component of the electric current,   
$J_{z}=e\langle\bar{\psi}\gamma^{3}\psi \rangle$, in view of the chiral magnetic effect.   
From eqs.~(\ref{n5_2}) and (\ref{CME_reg}), 
we find a simple relation between 
$J_z$ and the chirality imbalance $n_5$ 
in the LLL approximation ($n=0$) as
\begin{align}
  J_{z} &\simeq e\frac{|eB|}{2\pi}  \int \frac{d p_z}{2\pi} 
          f_{\Lambda}(p_z)[|\tilde{\phi}_{0, p_z}^{(+)}(t)|^2-|\tilde{\phi}_{0, p_z}^{(-)}(t)|^2]\nonumber \\
        &= e\frac{|eB|}{eB} n_5 \nonumber  \\
        & = \mathrm{sgn}(eB)e\, n_5 \label{eq:n5-and-jz-in-lll }
\end{align}
Eq.~(\ref{eq:n5-and-jz-in-lll }) tells us that 
$J_z$ is essentially proportional to $n_5$ in the limit of the strong magnetic field, where the use of the 
LLL approximation can be justified.  
The result agrees with one obtained in the previous work\cite{CME3},  
although the existence of the chiral chemical potential is assumed  in ref.~\cite{CME3}.  
Here, we recover eq.~(\ref{eq:n5-and-jz-in-lll })  by considering the massive fermion under  the external EM fields, 
without assuming the chirality asymmetric source.

\subsection{Chiral anomaly with the regularization}

We are also interested in the modification of the conservation law for the axial-vector current\cite{ABJ}.  
Neglecting the surface term from the current divergence, 
the chiral anomaly relation is given by
\begin{align}
\partial_t \int d^3 x \langle\bar{\psi}\gamma^0\gamma^5\psi\rangle 
=2im \int d^3 x \langle\bar{\psi} \gamma^5\psi\rangle + 
\frac{2 \alpha }{\pi} \int d^3 x {\bm E} \cdot {\bm B}  \, .
\label{anomly}
\end{align}   
Here, the second term of RHS is just an input of the model calculation in our case.  
On the other hand, we have already calculated the LHS and the 1st term of RHS individually.   Thus,  we can check  a consistency of our 
calculations with the point-split regularization.    

For the LHS of the chiral anomaly, we simply calculate the time-derivative of the chirality imbalance.   
If we used eq.~(\ref{n5}) for the chirality imbalance without invoking the momentum regularization, the time derivative would 
yield, 
\begin{align}
  &\partial_t \langle\bar{\psi}\gamma^0\gamma^5\psi\rangle \nonumber \\
&=2im \cdot   2[g_{n-1, p_y}^2(x)-g_{n, p_y}^2(x)]\cos\theta_n\mbox{Im}[\tilde{\phi}_{n, p_z}^{(+)}(t) \phi_{n, p_z}^{(-)}(t)]
\nonumber \\
&=2im\langle\bar{\psi}\gamma^5\psi\rangle \;\; .
\end{align}  
This is just a {\em classical} conservation law for the axial-vector current.

However, the gauge invariant regularization provides an additional time-dependent factor coming from  $\exp(-[p_z+eA_z(t)]^2/\Lambda^2)$, 
in the integrand of the chirality imbalance.  
Using the regularized result, eq.~(\ref{n5_2}), we obtain a modified conservation law as
\begin{align}
  &\partial_t \langle\bar{\psi}\gamma^0\gamma^5\psi\rangle =2im\langle\bar{\psi}\gamma^5\psi\rangle
    +\frac{2\alpha}{\pi}E_z(t)\, B\, F_{\Lambda}(t)
 \label{Anomary_rel}
\end{align}  
where 
\begin{align*}
  F_{\Lambda}(t)=\int^{\infty}_{-\infty} d p_z  \frac{p_z+eA_z(t)}{\Lambda_z^2}f_{\Lambda}(p_z)[|\tilde{\phi}_{0, p_z}^{(+)}(t)|^2-|\tilde{\phi}_{0, p_z}^{*(-)}(t)|^2] \; .
\end{align*}
This is the axial-vector current conservation law  in our framework.   
If the momentum cutoff is large enough, $\Lambda \gg m$, 
we obtain  a simple relation $\lim_{\Lambda\to \infty}F_{\Lambda}(t)=1$, which is explicitly shown 
in Appendix, 
and thus reproduce  the correct anomaly relation eq.~(\ref{anomly}).



\section{Time evolution of  the VEVs of the currents}
\label{section:Time evolution of  the VEVs of the currents}

In this section, we show numerical results for the time evolution of the chirality imbalance and CME in the vacuum.  
Here, we have three independent parameters of the model, magnitudes of the electric and magnetic fields, and the fermion mass, 
which are expressed in units of the electron mass $m_e= 0.5$MeV.    
We also need the parameter $\Lambda$ in the regularization function, and take $\Lambda = 30 m_e$, which is 
much larger than the fermion mass scale.

In our study, we calculate the VEVs of the vacuum under 
parallel constant magnetic and  the time-dependent Sauter Electric fields, whose magnitudes can be fixed 
independently.  
However, as far as we understand, 
the chirality imbalance  is well studied  by considering  the magnetic helicity density $h$ defined in eq.~(\ref{MH}).  
As we have discussed in the previous section, our calculation is fully consistent to the chiral anomaly relation.  
In the massless limit, it is simplified as  
\begin{align}
\partial_t \int d^3 x \langle\bar{\psi}\gamma^0\gamma^5\psi\rangle &\simeq
\frac{2 \alpha }{\pi} \int {\bm E} \cdot {\bm B} \nonumber \\
&= - \partial_t  \left( \frac{2 \alpha }{\pi} \int d^3 x {\bm A } \cdot {\bm B} \right)
\end{align}  
since we consider the time-independent magnetic  field.  
Integrand of the RHS is just magnetic helicity $h(t)$.  Hence, with our EM field, the chirality imbalance becomes 
\begin{align}
n_5 (t=\infty) = - \frac{2 \alpha} {\pi} h(t=\infty ) = e^2 B E \tau / \pi^2, 
\end{align}
which is true only for massless fermions.  
Nevertheless, it is convenient to express the chirality imbalance (and CME) in unit of the magnetic helicity, $ e^2 B E \tau / \pi^2$.

\begin{figure}[h]
  \centering
  \includegraphics[width=8.5cm]{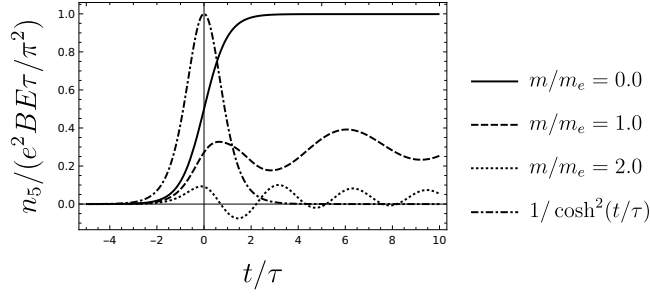}
  \caption{The time evolution of chirality imbalance $n_5$ $ \tau / \pi^2$.   
$eE/m_e^2=4.0, \tau m_e=0.5, eB/m_e^2=8.0$
  }
  \label{fig:01}
\end{figure}

In Fig.~\ref{fig:01}, we first show the chirality imbalance $n_5$ as a function $t$ with a shape of the Sauter electric field 
by  the dash-dotted curve.  
In the massless case (solid curve), $n_5$ increases by the electric field,  and approaches a finite value, 
$e^2 B E \tau / \pi^2$, at $t\to \infty$ even after $E$ field diminished.  
On the other hand, in the case of the finite fermion mass, the chirality imbalance consists of both  a constant part and 
an oscillating part at $t \to \infty$.  
When the mass is comparable with the magnitude of the electric field, $m^2 \sim (eE$),  
the chirality imbalance is largely suppressed as depicted by the dotted curve.  
Thus, we find that the average chirality imbalance is almost zero, if $m^2 > (eE$).   
We will relate these results with the fermion pair production from the vacuum in view of the Schwinger mechanism\cite{SchBog}.  

We also examine effects of the magnetic field on the chirality imbalance.  
If we increase the strength of the magnetic field, the magnitude of the chirality imbalance is also increased which is 
just proportional to the magnetic helicity.   However, the time-dependence of $n_5$ is never changed as expected.

\begin{figure}[h]
  \centering
  \includegraphics[width=8.5cm]{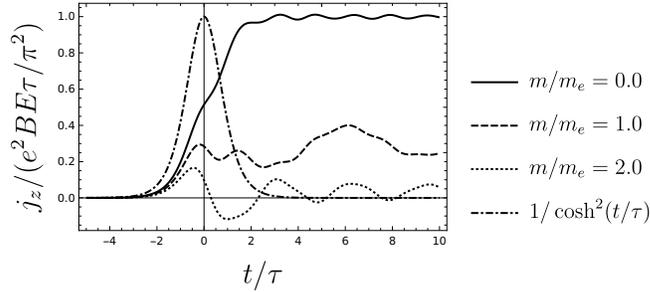}
  \caption{The time evolution of vector current density $j_z$  $ \tau m_e=0.5, eB/m_e^2=8.0$
  }
  \label{fig:02}
\end{figure}

We then show the vector current along the $z$-direction in Fig.~\ref{fig:02}  which could be understood as the chiral magnetic effect.  
Again, the vector current is shown in units of the magnetic helicity density.  
In the case of the massless limit, the vector current depicted by the solid curve consists of dominant constant part and 
tiny oscillating part, which is somewhat different from the behavior of the chirality imbalance $n_5$.  
This is because $n_5$ is solely determined by the lowest Landau level contribution, while the vector current gets 
contributions from higher Landau levels in eq.~(\ref{CME_reg}).   
The average CME current almost vanishes for  the small electric field  $m^2 > (eE)$, which is similar with the 
chirality imbalance.

From Fig.~\ref{fig:02}, for  $t/\tau\gg 1$ where there is no electric field, the CME current for the massless fermion is expressed as
 \begin{align}
j_z   \simeq \frac{e^2 B E \tau}{ \pi^2} = \frac{\alpha}{2\pi} B \left( 8E \tau \right) \;.
\label{CME_similar}
 \end{align}
The form of eq.(\ref{CME_similar}) is the same as eq.~(\ref{CME}) if we substitute $ 8E \tau$ for $\mu_5$.    
This crude identification is justified  only if $t$ is large enough compared with the time scale $\tau$ of the electric field in 
eq.~(\ref{Sauter}).

\begin{figure}[h]
  \centering
  \includegraphics[width=8.5cm]{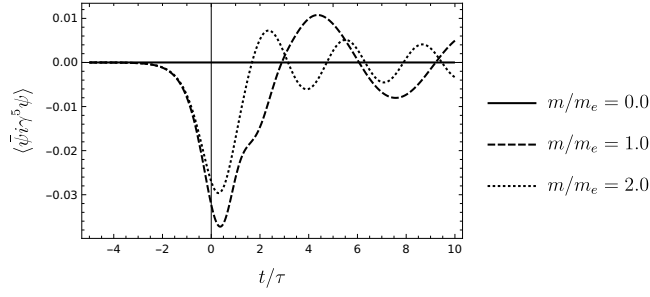}
  \caption{The time evolution of pseudoscalar density. $m_e=0.5 \mathrm{MeV}, eE/m_e^2=4.0, \tau m_e=0.5, eB/m_e^2=8.0$
  }
  \label{fig:07}
\end{figure}

For completeness, we also show the pseudoscalar density in Fig.\ref{fig:07} calculated with eq.~(\ref{PS_reg}).  
As expected from the chiral anomaly relation eq.~(\ref{Anomary_rel}),  
the pseudoscalar density is significant only at $t \sim 0$.  

\section{Relation to the fermion pair-prudcution}
\label{Relation to the fermion pair-prudcution}

In order to understand appearance  of the chirality imbalance from the vacuum, we relate it with the 
fermion pair-production\cite{SchBog,Tanji}.   
To do so, we try to find a relation between the "in-state" vacuum at $t \to -\infty$ and the "out-state" vacuum 
at $t \to \infty$.    
As discussed in eqs.~(\ref{eq:f-in-plus-asymptotic-free-field},\ref{eq:f-in-minus-asymptotic-free-field}), 
our original "in state" vacuum at $t \to -\infty$, $|0\rangle$, coincides with the free particle vacuum (although $B\neq 0$).  
However, due to the Sauter-electric field, the vacuum at $t \to \infty$, $|0\rangle_{\mbox{out}}$, is not the same as 
the original vacuum $|0\rangle$.   

To proceed calculations, we need  asymptotic forms of 
the eigenfunction  $\tilde{\phi}_{{n, p_z}}^{(+)}(t)$ at $t \to -\infty$ and at $t \to \infty$.   
For the "in-state", the eigenfunctions reduce to
 \begin{align}
&\tilde{\phi}_{n,p_z}^{(+)}(t) \propto \exp(-i\omega_{n,p_z}(0)t) ~~~(t\to -\infty)
    \label{eq:f-in-plus-asymptotic-free-field}\\
  &\tilde{\phi}_{n,p_z}^{(-)}(t) \propto \exp(+i\omega_{n,p_z}(0)t) ~~~(t\to -\infty)
    \label{eq:f-in-minus-asymptotic-free-field}
\end{align} 
which agree with positive/negative energy plane wave solutions with the energy $\pm \omega (0)$ 
defined eq.~(\ref{omega_def}).  
On the other hand,  with the help of the connection formula for the Gauss hypergeometric function, 
"out-state" eigenfunctions are rewritten as 
\begin{align}
 \tilde{\phi}_{{n, p_z}}^{(+)}(t)
  &=\alpha_{n, p_z}\tilde{\phi}_{\mathrm{out},{n, p_z}}^{(+)}(t)-\beta_{n, p_z}^{*}\tilde{\phi}_{\mathrm{out},{n, p_z}}^{(-)}(t)\\
  \tilde{\phi}_{{n, p_z}}^{(-)}(t)
  &=  \alpha_{n, p_z}^{*}\tilde{\phi}_{\mathrm{out}, {n, p_z}}^{(-)}(t) + \beta_{n, p_z}\tilde{\phi}_{\mathrm{out}, {n, p_z}}^{(+)}(t)
\end{align}
where 
\begin{align}
    \tilde{\phi}_{\mathrm{out},n, p_z}^{(+)}(t) &\equiv \sqrt{\frac{\omega(1)+[p_z+2eE\tau]}{2\omega(1)}}u^{-\frac{i\tau\omega(0)}{2}}(1-u)^{\frac{i\tau\omega(1)}{2}}\nonumber \\
&\times
F\left({a,b \atop 1+a+b-c};1-u\right)\\
  \tilde{\phi}_{\mathrm{out},n, p_z}^{(-)}(t) &\equiv \sqrt{\frac{\omega(1)-[p_z+2eE\tau]}{2\omega(1)}}u^{\frac{i\tau\omega(0)}{2}}(1-u)^{-\frac{i\tau\omega(1)}{2}}\nonumber \\
&\times F\left({1-a,1-b \atop 1+c-a-b};1-u\right)  \; .
\end{align}  
and
\begin{align}
  \alpha_{n, p_z}
  &  = \sqrt{\frac{\omega(0) + p_z}{\omega(0)}}\sqrt{\frac{\omega(1)}{\omega(1)+[p_z+2eE\tau]}} \frac{2i}{\tau[\omega(0)+\omega(1)-2eE\tau]}\nonumber\\
  & \hspace{2cm} ~~\times \frac{\Gamma(1-i\tau\omega(0))\Gamma(-i\tau\omega(1))}{\Gamma(-\frac{i\tau\omega(0)}{2}-\frac{i\tau\omega(1)}{2}-ieE\tau^2)\Gamma(-\frac{i\tau\omega(0)}{2}-\frac{i\tau\omega(1)}{2}+ieE\tau^2)}\\
  \beta_{n, p_z}
  & =\sqrt{\frac{\omega(0) + p_z}{\omega(0)}}\sqrt{\frac{\omega(1)}{\omega(1)-[p_z+2eE\tau]}}  \frac{2i}{\tau[\omega(0)-\omega(1)-2eE\tau]}\nonumber\\
  & \hspace{2cm}~~\times \frac{\Gamma(1+i\tau\omega(0))\Gamma(-i\tau\omega(1))}{\Gamma(\frac{i\tau\omega(0)}{2}-\frac{i\tau\omega(1)}{2}+ieE\tau^2)\Gamma(\frac{i\tau\omega(0)}{2}-\frac{i\tau\omega(1)}{2}-ieE\tau^2)} \; .
\end{align}
$\tilde{\phi}_{\mathrm{out},n, p_z}^{(+)}(t) $ and $\tilde{\phi}_{\mathrm{out},n, p_z}^{(-)}(t) $ are further simplified as
\begin{align}
  &\tilde{\phi}_{\mathrm{out},n, p_z}^{(+)}(t) \propto \exp(-i\omega(1)t) ~~~(t\to \infty)\\
  &\tilde{\phi}_{\mathrm{out},n, p_z}^{(-)}(t) \propto \exp(i\omega(1)t) ~~~(t\to \infty) \; . .
\end{align}
which are the free fermion wave functions with the energy $\omega (1)$.    
From these functions, we can construct the Bogoliubov transformation between "in-state" and "out-state"\cite{SchBog,Tanji}.  
We already introduced the annihilation operators and the vacuum for the "in-state".
\begin{align}
  \hat{b}_{s,\boldsymbol{p}}|0\rangle
  = 0, ~\hat{d}_{s,\boldsymbol{p}}|0\rangle=0~(\text{for all }s,\boldsymbol{p}),~~\langle 0|0\rangle=1 \; .
\end{align}
Similarly, we define the "out-state" vacuum with operators $\hat{b}^{\mathrm{(out)}}_{s,\boldsymbol{p}}, \hat{d}^{\mathrm{(out)}}_{s,\boldsymbol{p}}$, 
\begin{align}
  \hat{b}^{\mathrm{(out)}}_{s,\boldsymbol{p}}|0\rangle_{\mbox{out}}
  = 0, ~\hat{d}^{\mathrm{(out)}}_{s,\boldsymbol{p}}|0\rangle_{\mbox{out}}   =0~(\text{for all }s,\boldsymbol{p}),~~\langle 0|0\rangle=1 \; .
\end{align}
where operators $\hat{b}^{\mathrm{(out)}}_{s,\boldsymbol{p}}$, 
$\hat{d}^{\mathrm{(out)}}_{s,\boldsymbol{p}}$ are introduced 
as coefficients of $\tilde{\phi}_{\mathrm{out},n, p_z}^{(+)}(t) $ and $\tilde{\phi}_{\mathrm{out},n, p_z}^{(-)}(t) $ in the 
standard way.  
Thus, these operators are subject to the transformation, 
\begin{align}
  \begin{pmatrix}
    \hat{b}_{\mathbf{p,s}}^{\mathrm{(out)}}\\
    \hat{d}_{\mathbf{p,s}}^{\mathrm{(out)} \, \dagger}
  \end{pmatrix}
  =
  \begin{pmatrix}
    \alpha_{n,p_z} & \beta_{n,p_z}\\
    -\beta_{n,p_z}^{*} & \alpha_{n,p_z}^{*}
  \end{pmatrix}
  \begin{pmatrix}
    \hat{b}_{\mathbf{p,s}}\\
    \hat{d}_{\mathbf{p,s}}^\dagger
  \end{pmatrix}
\end{align}
where the Bogoliubov coefficients satisfy the unitary condition $ |\alpha_{n, p_z}|^2+|\beta_{n, p_z}|^2=1 $.
The expectation value of the number operator at 
$t = \infty$ between the original vacuum becomes
\begin{align} 
\langle 0 | \hat{b}_{\mathbf{p,s}}^{\mathrm{(out)} \, \dagger} \, \hat{b}_{\mathbf{p,s}}^{\mathrm{(out)}}   | 0 \rangle = 
|\beta_{n, p_z}|^2
\end{align}
which is understood as the probability to find a fermion produced by the electric field 
with the momentum ${n, p_z}$ at $t =\infty$\cite{SchBog,Tanji,Taya}.  
It is well-known that $|\beta_{n, p_z}|^2$ is significant only if the electric field is larger than the fermion mass square, 
$eE>m^2$, which means spontaneous creation of fermion pairs  from the vacuum under the strong electric field.  
Thus, we naively expect the chirality imbalance may emerge for $eE\gg m^2$.

Using these results, one can express the VEVs of the vacuum at $t=\infty$ in terms of the Bogoliubov coefficients.  
%
For example,  the chirality imbalance $n_5$ at $t=\infty$ is calculated as 
\begin{align}
  n_5|_{t=\infty}&
= \frac{eB}{2\pi}\int \frac{d p_z}{2\pi}  f_{\Lambda}(p_z) \left[ -2|\beta_{0, p_z}|^2\frac{p_z+2eE\tau}{\omega(1)} \right. 
\nonumber\\
    & \hspace{1.5cm} \left.  -2\frac{m} {\omega(1)} \mbox{Re}[\alpha_{0, p_z}\beta_{0, p_z}\mathrm{e}^{-2i\omega(1)t}] \right]
\label{n5_Bog}
\end{align}
where the regularization function at $t=\infty$ is
\begin{align}
f_\Lambda ( p_z) = \exp[-\varepsilon{(p_z+2eE \tau )^2/\Lambda ^2 }] 
\end{align}
The first term is independent of time, and simply proportional to 
$|\beta_{0, p_z}|^2$ which is the probability to find a produced particle in the lowest Landau level with $p_z$.  
On the other hand, the second term is proportional to the mass, 
and is somehow interpreted as the "interference" term.

At first sight, $n_5$ is simply determined by  the magnitude of $|\beta_{0, p_z}|^2$.  
However, existence of the chirality imbalance strongly depends on details of the integration over $p_z$ in eq.~(\ref{n5_Bog}), 
which is sensitive to a parameter $\tau$, the time scale of the electric field in eq.~(\ref{Sauter}).  
We will discuss how the non-zero $n_5$ appears in some detail.   
In the massless limit, the first term of eq.~(\ref{n5_Bog}), which we call $n_5 ^{(0)}$, becomes
\begin{align}	
  n_5 ^{(0)}  
&= \frac{eB}{2\pi}\int \frac{d p_z}{2\pi}  f_{\Lambda}(p_z) \left[ -2|\beta_{0, p_z}|^2\frac{p_z+2eE\tau}{\sqrt{(p_z+2eE\tau)}
^2 + m^2}  \right] \nonumber\\
&\to 
\frac{eB}{2\pi}\int \frac{d p_z}{2\pi}  f_{\Lambda}(p_z) \left[ -2\, \mbox{sgn}[{p_z+2eE\tau} ]  
\, |\beta_{0, p_z}|^2 \,  \right] \; .
\label{dis_n5}
\end{align}

In the presence of the uniform magnetic field,  
all the fermions move along the $z$-direction, and the spin of the fermions in the lowest Landau level, which can contribute to 
$ |\beta_{0, p_z}| $, is  parallel to the z-direction.  
Hence, the fermions with positive canonical momenta,  
$p_z+2eE\tau >0$, carry the right-handed chirality, while 
those with $p_z+2eE\tau <0$ are left-handed.    
If  the electric field were zero, the imbalance $n_5 ^{(0)} $ would  vanish,  
because of a cancellation between contributions from $p_z>0$ and $p_z<0$ fermions by virtue of the  
symmetrical $p_z$ distribution of the pair-production probability $\beta_{0, p_z}$.  
However, non-zero electric field induces an asymmetry between momentum distributions of 
right-  and left-handed fermions in both the sign function $\mbox{sgn}[{p_z+2eE\tau} ] $ and the 
regularization function $f_\Lambda$, 
which indeed generates the chirality imbalance in this model.

\begin{figure}[h]
  \centering
  \includegraphics[width=8.5cm]{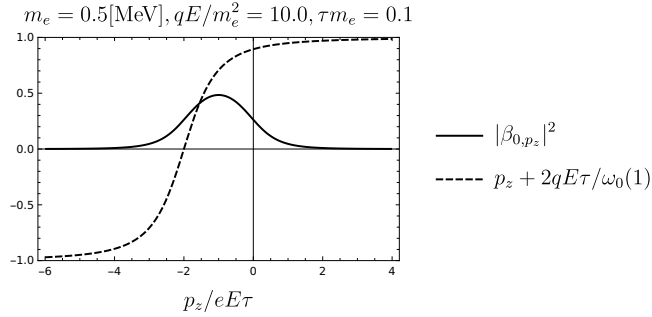}
  \caption{$p_z$ distribution of the pair-production probability $|\beta|^2$ and 
 the signfunction of $p_z+2eE\tau/\omega(1)$. }  
  \label{fig:05}
\end{figure}

To study  $n_5 ^{(0)} $ in the case of the finite mass, we show 
$|\beta_{0, p_z}|^2$ and   $(p_z+2eE\tau) /\omega(1)$ in Fig.~\ref{fig:05}, where 
$(p_z+2eE\tau) /\omega(1)$ is no longer the sign function.  
The pair-creation probability $|\beta_{0, p_z}|^2$ peaked at $p_z = -eE \tau$, whereas $(p_z+2eE\tau) /\omega(1)$ 
changes its sign 
at  $p_z = -2eE \tau$.    Hence, if $\tau$ is very small ($\sim 0)$,  the integration over $p_z$ is negligible due to a cancellation, 
and thus the resulting chirality imbalance almost vanishes.   

\begin{figure}[h]
  \centering
  \includegraphics[width=8cm]{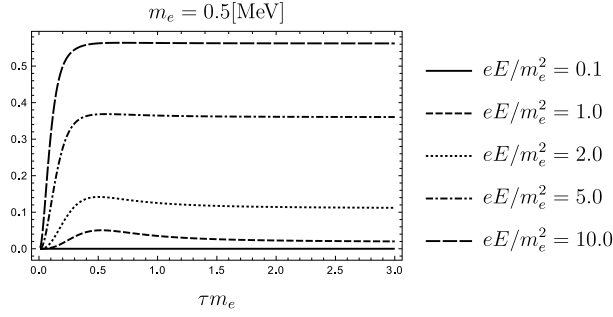}
  \caption{$\tau$ dependence of $n_5 ^{(0)}$ for several values of $eE$.   }
  \label{fig:04}
\end{figure}

\begin{figure}[h]
  \centering
  \includegraphics[width=8cm]{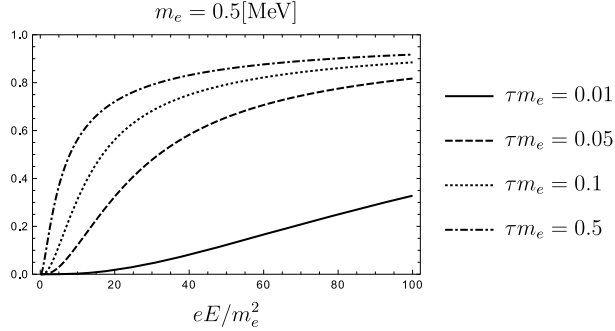}
  \caption{$eE$ dependence of $n_5 ^{(0)}$  }
  \label{fig:03}
\end{figure}

For completeness, we show explicit $\tau$  dependence of the results.  
We first show  the chirality imbalance as a function of $\tau m_e$ in Fig.~\ref{fig:04} for several values of $eE$.  
If $eE< m^2$, the chirality imbalance is almost zero, because the production of the fermion pairs is forbidden.  
The large electric field simply gives  the larger chirality imbalance.    
However, if the time scale $\tau$ is quite small, $\tau \ll 1/m_e$, situation becomes different.   
In Fig.~\ref{fig:03},  we show $n_5^{(0)}$ for several values of $\tau m_e$.  We find that, even if the magnitude of the 
electric field is large enough, 
$n_5^{(0)}$ is very small for $\tau m_e < 0.01$.  This is because the small $\tau$ cannot provide enough 
asymmetry in the integrad of eq.~(\ref{dis_n5}).

Similar argument holds for the chiral magnetic effect,  the $z$-component of vector current at $t = \infty$.  
We can write  the CME current in  terms of  Bogoliubov coeffcients as, 
\begin{align}
  \langle\bar{\psi}\gamma^3\psi\rangle|_{t=\infty}
  &= -\frac{|eB|}{2\pi}\int \frac{d p_z}{2\pi}  f_{\Lambda}(p_z)
  \sum_{n=0}^{\infty}\alpha_n \left[-2|\beta_{n, p_z}|^2\frac{p_z+2eE\tau}{\omega(1)} \right. \nonumber\\
    &\left. \hspace{1cm} -2\frac{\sqrt{m^2+2eBn}}{\omega(1)}\mbox{Re}[\alpha_{n, p_z}\beta_{n, p_z}\mathrm{e}^{-2i\omega(1)t}]
\right], 
    \label{dis_CME}
\end{align}
which is  similar with one of the chirality imbalance.  
The first term is independent of time and essentially given by a product of $|\beta_{0, p_z}|^2$ and 
 $(p_z+2eE\tau) /\omega(1)$, which is interpreted as 
the $z$-component of relativistic velocity of particles.     
Hence, this term is understood as a classical analogue of the electric current of the $z$-component carried by the produced 
fermions.  
Note that the second oscillating term is non-zero even in the massless limit.

\section{Summary and Disscussions}
\label{label:Summary and Disscussions}

We have studied the chirality imbalance of the vacuum under the time-independent magnetic  field and the Sauter-type 
pulsed electric field.   In particular, we have focused on the time evolution of the chirality imbalance and the chiral magnetic effect 
from $t=-\infty$ to $t = \infty$.  
Solving the squared Dirac equation with the EM field, we have constructed the quantized fermion field and the
 vacuum at $t=-\infty$.   
Then, we have calculated the vacuum expectation values of various fermion current operators including $n_5$ and CME in terms of the point-split 
regularization.  Use of the gauge invariant regularization method is important in our study, because the VEVs diverge 
by the momentum integration.   
Subtle cancellation between positive and negative energy states provides non-zero contribution to CME.  
We note that calculated VEVs are finite at $t=\infty$, and differ from the case with the  constant electric and magnetic fields, where 
the several VEVs become infinite at $t\to \infty$\cite{Warringa:2012bq}.   
In addition, we have  found expressions for the VEVs of other bilinear operators, e.g.~ 
$\langle \bar \psi \gamma_3\gamma_5\psi \rangle = 0$, whereas  $\langle \bar \psi \sigma_{12}\psi\rangle \neq 0$.

We have shown the time evolution of the chirality imbalance and CME current.     
The resulting chirality imbalance  is finite at $t=\infty$ where the Sauter electric field is already turned off.  
We have also demonstrated that a part of the chirality imbalance consists of the time-oscillating contribution, which 
is proportional to the fermion mass.    
The CME current also consists of
the dominant time-independent part and the oscillating part, which is similar with the chirality imbalance.  
We have also discussed a connection between the fermion pair-creation by the electric field 
and the chirality imbalance.  As we have obtained in eqs.(\ref{dis_n5},\ref{dis_CME}), there are simple physical interpretations for the 
generation of $n_5$ and CME.

The magnitudes of $n_5$ and CME in this model are essentially determined by the following conditions;  
\begin{enumerate}
\item enough magnitude of the  electric field which is much larger than the fermion mass scale. 
\item enough asymmetric $p_z$ distribution of the produced fermions (in the integrand of the chirality imbalance).  
\end{enumerate}
Asymmetries of the pair-production rate between $p_z>0$ and $p_z<0$ particles are important to produce the chirality imbalance, and 
may depend on details of the external electromagnetic fields.  
In fact, magnitudes of the chirality imbalance for the massive fermion change largely if we change the time dependence of the electric field by using 
the Gaussian packet formalism\cite{AoiSuzuki2}.

Although we consider the external electromagnetic fields in this work, it may be important to include "back-reaction" of the 
external field.  It is possible to include back-reaction effects on the electric field within this framework\cite{SchBog}.  
Work along this line is under consideration.

\vspace{2cm}

\noindent
{\bf Acknowledgments}

We acknowledge Prof.~A.~Suzuki for  useful and critical comments.    
We also thank all the members of the quark/hadron group in Tokyo University of
Science for useful conversations.   
K.S. thanks Dr.~S.~Saitoh for contributions on early stage of this work.  

\newpage

\noindent\noindent
{\large \bf Appendix : Anomaly relation}

We will provide a proof of the anomaly equation, eq.(\ref{Anomary_rel}), 
\begin{align}
  &\partial_t \langle\bar{\psi}\gamma^0\gamma^5\psi\rangle =2im\langle\bar{\psi}\gamma^5\psi\rangle
    +\frac{2\alpha}{\pi}E_z(t)\, B\, F_{\Lambda}(t)
\end{align}  
where 
\begin{align*}
F_{\Lambda}(t)=\int^{\infty}_{-\infty} d p_z  \frac{p_z+eA_z(t)}{\Lambda_z^2}\exp(-\frac{(p_z+eA(t))^2}{\Lambda^2})[|\tilde{\phi}_{0, p_z}^{(+)}(t)|^2-|\tilde{\phi}_{0, p_z}^{*(-)}(t)|^2] 
\to 1 ~~(\Lambda \to \infty)
\end{align*}
Using the solutions of the Dirac equation for the lowest Landau level, $\tilde{\phi}_{0, p_z}$, 
we first rewrite the integrand of $F_\Lambda(t)$ as
\begin{align}
  |\tilde{\phi}_{0, p_z}^{(+)}(t)|^2-|\tilde{\phi}_{0, p_z}^{*(-)}(t)|^2  = \frac{p_z +eA(t)}{\sqrt{m^2+(p_z -eA(t))^2}}+ G(p_z,t) \;.
\label{App-2nd}
\end{align}
where the first term gives a finite contribution as $p_z\to \infty$, while 
the second term,  $G(p_z,t)$, is  a rapidly decreasing function, $\lim_{ |p_z|\to \infty} |G(p_z,t)|\to 0$.   
We decompose the integrand of the first term using 
\begin{align}
\frac{p_z^2}{\sqrt{m^2+p_z^2}}=|p_z| +|p_z| \, H(\frac{p_z^2}{m^2})\; , 
\end{align}
where a function $H(p_z,t)$ satisfies $\lim_{ p_z\to \infty} H(p_z/m,t)\to 0$.   Thus, 
we rewrite the first term  as 
\begin{align*}
&\int^{\infty}_{-\infty} d p_z  \frac{p_z+eA_z(t)}{\Lambda^2}\exp(-\frac{(p_z+eA(t))^2}{\Lambda^2})
\frac{p_z +eA(t)}{\sqrt{m^2+(p_z -eA(t))^2}}\nonumber\\
=& \frac{2}{\Lambda^2}\int^{\infty}_{0} d p_z  \, |p_z|\exp(-\frac{p_z^2}{\Lambda^2})
+\frac{2}{\Lambda^2}\int^{\infty}_{0} d p_z \, |p_z|\, H(\frac{p_z^2}{m^2})\,\exp(-\frac{p_z^2}{\Lambda^2})\\
 =& 1+ \frac{2m^2}{\Lambda^2}\int^{\infty}_{0} du\,  H(u)\exp(-\frac{m^2}{\Lambda^2}u)\\
  \to& 1
\end{align*}
as $\Lambda\to \infty$, because of the fact $H (\frac{p_z^2}{m^2}) \to 0$ ($p_z\to\infty$).

On the other hand, the second term of eq.(\ref{App-2nd}) can be shown to vanish in the similar way.  
In our model with the Sauter electric field, 
$G(p_z,t)$, decreases rapidly for  $|p_z| \gg eE\tau$ independent of time $t$, as shown in 
Fig.~\ref{fig:05}.   
Hence, in the limit $\Lambda \to \infty$,  i.e.~$\Lambda \gg eE\tau$, 
the integal of  the second term is independent of $\Lambda$,  and proportional to 
$(eE\tau)^2$ by the dimensional analysis.  (If the fermion mass is comparable with  
$(eE)^{1/2}$, the argument should be modified.  However, the essential result is not changed for $\Lambda \gg (eE)^{1/2}$ or $m$.)  
It leads to 
\begin{align*}
&\int^{\infty}_{-\infty} d p_z  \frac{p_z+eA_z(t)}{\Lambda_z^2}\exp(-\frac{(p_z+eA(t))^2}{\Lambda^2})G(p_z,t)
\\=&\frac{(eE\tau)}{\Lambda^2}\times \mbox{($\Lambda$-independent constant)}
\to 0
\end{align*}
Thus, we recover the correct anomaly relation in eq.~(\ref{Anomary_rel}).


\renewcommand{\theequation}{A.\arabic{equation} }
\setcounter{equation}{0}


\end{document}